\documentclass[aps,prd,preprint,superscriptaddress]{revtex4}

\usepackage{graphicx}

\def\bwt{\begin{widetext}}
\def\ewt{\end{widetext}}
\def\be{\begin{equation}}
\def\ee{\end{equation}}
\def\bea{\begin{eqnarray}}
\def\eea{\end{eqnarray}}
\def\bean{\begin{eqnarray*}}
\def\eean{\end{eqnarray*}}
\def\bary{\begin{array}}
\def\eary{\end{array}}
\def\bit{\begin{itemize}}
\def\eit{\end{itemize}}

\def\su5u1{SU(5) \times U(1)}
\def\fsu5u1{SU(5) \times U(1)'}
\def\so10{SO(10)}
\def\sq20{SO(10) \times SO(10)}

\begin{document}

\title{Non-Canonical Gauge Coupling Unification in High-Scale Supersymmetry Breaking}

\author{V. Barger}
\affiliation{Department of Physics, University of Wisconsin, 
Madison, WI 53706, USA}

\author{Jing Jiang}
\affiliation{Institute of Theoretical Science, University of Oregon, 
Eugene, OR 97403, USA}

\author{Paul Langacker}
\affiliation{Department of Physics and Astronomy,
University of Pennsylvania, Philadelphia, PA 19104-6396, USA}

\author{Tianjun Li}
\affiliation{School of Natural Sciences, Institute for Advanced Study,
  Einstein Drive, Princeton, NJ 08540, USA}

\date{\today}


\begin{abstract}

The string landscape suggests that the supersymmetry breaking scale
can be high, and then the simplest low energy effective theory is 
the Standard Model (SM). Considering grand unification 
scale supersymmetry breaking,
we show that gauge coupling unification can be
achieved at about $10^{16-17}$ GeV in the SM 
with suitable normalizations of the $U(1)_Y$, 
and we predict that the Higgs mass range is
127 GeV to 165 GeV, with the precise value strongly correlated with
the top quark mass $m_t$ and $SU(3)_C$ gauge coupling.
For example, if $m_t=178\pm1$ GeV, the Higgs boson mass is predicted to be
between $141$ GeV and $154$ GeV.
We also point out that gauge coupling unification in the Minimal
Supersymmetric Standard Model (MSSM)  
does not imply the canonical $U(1)_Y$ normalization.
In addition, we  present 7-dimensional 
orbifold grand unified theories (GUTs) in which such normalizations for the
$U(1)_Y$ and charge quantization can be realized.
The supersymmetry can be broken at the grand unification scale 
by the Scherk--Schwarz mechanism. We briefly comment on
a non-canonical $U(1)_Y$ normalization due to the
brane localized gauge kinetic terms in orbifold GUTs.

\end{abstract}


\preprint{MADPH-05-1421, OITS-767, UPR-1113-T, hep-ph/0504093}

\maketitle

\section{Introduction}

The great mystery in particle physics is the cosmological
constant problem: why is the cosmological constant $\Lambda_{\rm CC}$ so tiny
compared to the Planck scale $M_{\rm Pl}$ or the string scale, {\it i.e.},
$\Lambda_{\rm CC} \sim 10^{-122} M_{\rm Pl}^4$? 
There is no known symmetry in string theory that constrains
the cosmological constant to be zero.
Another major puzzle is the gauge hierarchy problem. 
In the Standard Model (SM), radiative corrections to the Higgs boson 
(or a scalar in general) mass is quadratically
dependent on the UV cutoff scale, and its mass is unprotected
by any chiral or gauge symmetry. Thus, the natural Higgs mass is of
the order of the UV cutoff scale rather than the weak scale.  Plausibly
the UV cutoff scale should be around the Planck scale or string 
scale. The many orders of magnitude difference between the UV cutoff
scale and the weak scale is the gauge hierarchy
problem.  A well known solution to this problem is supersymmetry.
 However,  supersymmetry can ameliorate but does not solve the 
cosmological constant problem.

In string models with flux compactifications
there exists an enormous ``landscape'' for long-lived
metastable string/M theory vacua where the
moduli can be stabilized and supersymmetry may be broken~\cite{String}. 
In particular,  applying the ``weak anthropic principle''~\cite{Weinberg}, the 
string landscape proposal may 
provide the first concrete explanation of the very tiny value of the cosmological
constant, which can take only discrete values, and it may also address the gauge 
hierarchy problem.  Notably, the supersymmetry breaking scale 
can be high if there exist many supersymmetry breaking parameters or many hidden
sectors~\cite{HSUSY,NASD}.  Although there is no definite conclusion whether
the string landscape predicts high-scale or TeV-scale supersymmetry 
breaking~\cite{HSUSY},
it is interesting to study models with high-scale supersymmetry
breaking as we await the turn on of the Large Hadron Collider (LHC)~\cite{NASD,Barger:2004sf}.

If supersymmetry is indeed broken at a high scale, the breaking scale can range
from 1 TeV to the string scale. 
Three representative choices for the supersymmetry breaking scale 
can be considered\cite{Barger:2004sf}: 
(1)~the string scale, (2)~an intermediate scale, and (3)~the TeV scale. Because
 TeV-scale supersymmetry has been studied extensively during the 
last two decades,  we do not consider it here.
We emphasize that for string-scale and intermediate-scale 
supersymmetry breakings,
most of the problems associated with low energy supersymmetric models, for example,  
excessive flavor and CP violations, dimension-5 fast proton decay and the stringent 
constraints on the lightest CP even neutral 
Higgs boson mass, are solved automatically.

For intermediate-scale supersymmetry breaking, 
Arkani-Hamed and Dimopoulos proposed the
split supersymmetry scenario where the scalars (squarks, sleptons and one
combination of the scalar Higgs doublets) have masses at an
intermediate scale, while the fermions (gauginos and Higgsinos) and the other
combination of the scalar Higgs doublets are at the TeV
scale~\cite{NASD}. Gauge coupling unification is preserved and the lightest
neutralino can still be a dark matter candidate. 
The realization and phenomenological consequences of split supersymmetry
have been studied in Refs.~\cite{ssphen1,ssphen2,ssphen3,ssphen4,ssphen5,ssphen6}.

However, unlike the cosmological constant problem and the gauge
hierarchy problem, the strong CP problem is still a challenge for naturalness
 in the string landscape~\cite{Donoghue:2003vs}.  In the Standard Model, the
$\overline{\theta} $ parameter is a dimensionless coupling constant which is
infinitely renormalized by radiative corrections.  There is no theoretical
reason for $\overline{\theta} $ as small as $10^{-9}$ required by the
experimental bound on the electric dipole moment of the neutron~\cite{review,
  pdg}.  There is also no known anthropic constraint on the value of
$\overline{\theta} $, {\it i.e.}, $\overline{\theta} $ may be a random variable
with a roughly uniform distribution in the string
landscape~\cite{Donoghue:2003vs}.
In addition,  from flux-induced
supersymmetry breaking in Type IIB orientifolds~\cite{Camara:2004jj}, 
 supersymmetry breaking soft masses are all approximately of the same order
in general.  Supersymmetric axion models with an approximately universal 
  intermediate-scale ($\sim 10^{11}$~GeV) supersymmetry breaking were
proposed in Ref.~\cite{Barger:2004sf},
 where the strong CP problem is solved by the 
well-known Peccei--Quinn mechanism~\cite{PQ}.
The global PQ symmetry is protected against quantum 
  gravitational violation by considering the gauged discrete $Z_N$ PQ 
  symmetry \cite{Babu:2002ic}, which can be
embedded in an anomalous $U(1)_A$ gauge symmetry in string constructions
where the anomalies can be cancelled by the Green--Schwarz
mechanism~\cite{MGJS}. In these models, the axion can be a cold dark matter 
  candidate, and the intermediate supersymmetry breaking scale is 
  directly related to the PQ symmetry breaking scale. Gauge coupling 
  unification can be achieved at about $2.7\times 10^{16}$~GeV due to 
 additional SM vector-like fields at the intermediate scale, and
  the Higgs mass range is from 130~GeV to 160~GeV \cite{Barger:2004sf}.

For string-scale supersymmetry breaking, 
 axion models in which gauge coupling unification 
can be realized by introducing SM vector-like fermions were
discussed in Ref.~\cite{Barger:2004sf}.
Then, we proposed the SM as a low energy effective theory where
gauge coupling unification can be achieved by choosing
suitable $U(1)_Y$ normalizations~\cite{Barger:2005gn}.

In this paper, we consider an approximately universal
 grand unification scale or string-scale supersymmetry breaking \cite{Barger:2004sf}.
To solve the strong CP problem in the string landscape, 
we adopt the PQ mechanism~\cite{PQ} with
the axion as the cold dark matter candidate.
Because the supersymmetry breaking scale is around the grand unification
scale or the string scale, the minimal model at low energy 
is the Standard Model. The SM explains existing experimental data very well,
 including electroweak precision tests. In addition, 
it is easy to incorporate aspects of physics beyond the SM through small 
variations, for example,  dark matter, dark energy,
atmospheric and solar neutrino oscillations,
 baryon asymmetry and inflation~\cite{Davoudiasl:2004be}. 
The SM fermion masses and mixings can be
explained via the Froggatt-Nielsen mechanism~\cite{FN}.
However, there are still some limitations of the SM,
for example, the lack of explanation of gauge coupling unification and charge
quantization.

Charge quantization can easily be realized by
embedding the SM into a grand unified theory (GUT). 
Anticipating that the Higgs particle could be the only
new physics observed at the LHC, thus confirming the SM 
as the low energy effective theory, it is appropriate to 
reconsider gauge coupling unification in the SM.
However, it is well known that
gauge coupling unification cannot be achieved in the SM with
 the canonical normalization of the $U(1)_Y$ hypercharge
interaction, {\it i.e.}, the Georgi-Glashow $SU(5)$ 
normalization~\cite{Langacker:1991an}.
Gauge coupling unification can be achieved in the SM
by introducing extra  multiplets
between the weak and GUT scales~\cite{Frampton:1983sh}
or large threshold corrections~\cite{Calmet:2004ck},
which generically introduce more fine-tuning into the theory.
To avoid proton decay induced by
 dimension-6 operators via heavy gauge boson exchanges,
 the gauge coupling unification scale is constrained to 
be higher than about $5\times 10^{15}$ GeV.

We shall reexamine gauge coupling unification in the SM \cite{Barger:2005gn}.
The gauge couplings for $SU(3)_C$ and $SU(2)_L$ are unified at
about $10^{16-17}$ GeV, and the gauge coupling for the $U(1)_Y$
at that scale depends on its normalization.
If we choose a suitable normalization
of the $U(1)_Y$, the three gauge couplings for $SU(3)_C$, $SU(2)_L$
and $U(1)_Y$ can in fact be unified at about $10^{16-17}$ GeV,
and then there is no proton decay problem via dimension-6 operators.
Thus, the key question is:
{\it  is the canonical normalization for $U(1)_Y$ unique?}

For a 4-dimensional (4D) GUT with a simple group, 
 the canonical $U(1)_Y$ normalization is the only possibility, assuming that
 the SM fermions form complete multiplets
under the GUT group. However, the $U(1)_Y$
normalization need not be canonical in string model 
building \cite{Dienes:1996du,Blumenhagen:2005mu},
 orbifold GUTs \cite{Orbifold,Li:2001tx}
and their deconstruction \cite{Arkani-Hamed:2001ca}, and in 4D GUTs
with product gauge groups. We discuss these possibilities below:

(1)  In weakly coupled heterotic string theory, the gauge and
gravitational couplings always automatically unify at tree
level to form one dimensionless string coupling 
constant $g_{\rm string}$ \cite{Dienes:1996du}
\begin{eqnarray}
k_Y g_Y^2 = k_2 g_2^2 = k_3 g_3^2 = 8 \pi {{G_N}\over {\alpha'}}
= g_{\rm string}^2 ~,
\end{eqnarray}
where $g_Y$, $g_2$, and $g_3$ are the gauge couplings for
the $U(1)_Y$, $SU(2)_L$, and $SU(3)_C$ gauge groups, respectively;
$G_N$ is the gravitational coupling;
and $\alpha'$ is the string tension.
Here, $k_Y$, $k_2$ and $k_3$ are the levels of the corresponding
Kac-Moody algebras; $k_2$ and $k_3$ are positive integers while
$k_Y$ is in general a rational number \cite{Dienes:1996du}.

(2) In intersecting D-brane model building on Type II orientifolds,
the normalization for the $U(1)_Y$ (and also other gauge factors) 
is not canonical in general \cite{Blumenhagen:2005mu}.

(3) In orbifold GUTs \cite{Orbifold}, 
only the SM or SM-like gauge symmetry should be preserved on
the 3-branes at the fixed points  \cite{Li:2001tx}. Then
the SM fermions, which can be localized on 
the  3-brane at a fixed point, need not form 
complete multiplets under the GUT group. Thus, the $U(1)_Y$
normalization need not be canonical.
This statement also holds for the deconstruction of
orbifold GUTs \cite{Arkani-Hamed:2001ca} and for
 4D GUTs with product gauge groups.

In this paper we shall assume that at the GUT or string
scale, the gauge couplings in the SM satisfy
\begin{eqnarray}
g_1 = g_2= g_3 ~,
\end{eqnarray}
where $g_1^2 \equiv k_Y g_Y^2$, in which $k_Y=5/3$ for  
canonical normalization.
We show that gauge coupling unification in the SM can be
achieved at about $10^{16-17}$ GeV for $k_Y=4/3$, 5/4, 32/25.
Especially for $k_Y=4/3$, gauge coupling unification in the SM
is well satisfied at two loop order. 
In addition, with GUT scale supersymmetry 
breaking, we predict that the Higgs mass is in the range 127 GeV to 165 GeV
when the top quark mass is varied within its $2\sigma$ experimental range
and  the $SU(3)_C$ gauge coupling within its $1\sigma$ range.
The top quark mass can be measured
to about $1$ GeV accuracy at the LHC~\cite{Beneke:2000hk}.  
Assuming this accuracy and its
central value of $178$ GeV, the Higgs boson mass is predicted to be
between $141$ GeV and $154$ GeV.
Moreover, we point out that gauge coupling unification in the 
Minimal Supersymmetric Standard Model (MSSM) 
does not require $k_Y=5/3$, and we show that 
gauge coupling unification in the MSSM 
can  be achieved to the same degree by the choice $k_Y=7/4$.

Furthermore, on the space-time $M^4\times T^2/Z_6 \times S^1/Z_2$,
where $M^4$ is the 4D Minkowski space-time,
we construct a 7-dimensional (7D) $SU(6)$ model with $k_Y=4/3$
and 7D $SU(7)$ models with $k_Y=5/4$ and 32/25.
In these models, the $SU(6)$ and $SU(7)$ gauge symmetries
can be broken down to the SM-like gauge symmetries via orbifold
projections and be further broken down to the SM gauge symmetry by
the Higgs mechanism. A right-handed top quark in the $SU(6)$ model
and one pair of Higgs doublets in the $SU(7)$ models can
be obtained from the zero modes of the bulk vector multiplet, and their
hypercharges are determined from the constructions.
Then, charge quantization can be achieved from the anomaly free
conditions and the gauge invariance
of the Yukawa couplings. 
The extra $U(1)$ gauge symmetries can be considered
as  flavour symmetries, and the SM fermion masses and
mixings may be explained naturally via the Froggatt-Nielsen mechanism \cite{FN}.
The supersymmetry can be broken at the GUT scale 
by the Scherk--Schwarz mechanism \cite{Scherk:1978ta}.
We also briefly present a 7D orbifold $SU(8)$ model
with $k_Y=7/4$ and charge quantization, and comment on
the non-canonical $U(1)_Y$ normalization due to the
brane localized gauge kinetic terms in orbifold GUTs.

This paper is organized as follows: 
in Section II we study gauge coupling unification and the Higgs boson mass.
We discuss the 7D orbifold GUTs in Section III.
Discussion and conclusions are presented in Section IV. 
In Appendix A we give the relevant renormalization group equations.

\section{Gauge Coupling Unification and Higgs Boson Mass}

\subsection{Gauge Coupling Unification}

We define $\alpha_i=g_i^2/4\pi$ and denote the $Z$ boson
mass as $M_Z$.
In the following, we choose the top quark pole mass 
$m_t = 178.0\pm 4.3 $ GeV~\cite{Azzi:2004rc}, the strong coupling constant
$\alpha_3(M_Z) = 0.1182 \pm 0.0027$~\cite{Bethke:2004uy}, and the
fine structure constant $\alpha_{EM}$, weak mixing angle $\theta_W$ and 
Higgs vacuum expectation value $v$ at $M_Z$ to be~\cite{Eidelman:2004wy}
\bea
&&\alpha^{-1}_{EM}(M_Z) = 128.91 \pm 0.02\,, \nonumber \\
&&\sin^2\theta_W(M_Z) = 0.23120 \pm 0.00015\,, \nonumber \\
&&v = 174.10\,{\rm GeV}\,.
\eea

We first examine the one-loop running of the gauge couplings.
The one-loop renormalization group equations (RGEs) in the SM are
\begin{eqnarray}
(4\pi)^2\frac{d}{dt}~ g_i &=& b_i g_i^3~,~\,
\label{SMgauge}
\end{eqnarray}
where $t=\ln  \mu$ with $ \mu$ being the renormalization scale, and
\begin{eqnarray}
b\equiv (b_1, b_2, b_3)=\left(\frac{41}{6 k_Y},-\frac{19}{6},-7\right)~.~\,
\label{SMbi}
\end{eqnarray}
We consider the SM with $k_Y=4/3$, 5/4, 32/25 and 
the canonical 5/3. In addition, we consider
the extension of the SM with two Higgs doublets (2HD) with $b=(7/k_Y, -3, -7)$ 
and $k_Y=4/3$, and the MSSM with $b=(11/k_Y, 1, -3)$.
For the MSSM, we assume a supersymmetry breaking scale of 300 GeV
for scenario I (MSSM I), and an effective 
supersymmetry breaking scale of 50 GeV to include the threshold corrections
due to the mass differences between the squarks and sleptons for
scenario II (MSSM II) \cite{Langacker:1992rq}. 
For the MSSM I and MSSM II, the 
 $U(1)_Y$ normalization can be the canonical $k_Y=5/3$, or the alternative
$k_Y=7/4$. For different $k_Y$,
the initial values of $\alpha_1(M_Z)$ are normalized as $\alpha_1(M_Z)
= k_Y \alpha_Y$, where $\alpha_Y = \alpha_{EM}(M_Z)/\cos^2\theta_W(M_Z)$.

We use $M_{U}$ to denote the unification scale where $\alpha_2$ and
 $\alpha_3$ intersect in the RGE evolutions.  There is a sizable uncertainty
 associated with the $\alpha_3(M_Z)$ measurement.  To consider the
effects of the $\alpha_3(M_Z)$ uncertainty, we also use $\alpha_3 - \delta \alpha_3$
 and $\alpha_3 + \delta \alpha_3$ as the initial values for the RGE
 evolutions, whose corresponding unification scales with $\alpha_2$ are called
 $M_{U-}$ and $M_{U+}$, respectively.  Simple relative differences
 for the gauge couplings at the unification scale are defined as
\begin{eqnarray}
\Delta ~=~ {{|\alpha_1^{-1}(M_{U}) -
 \alpha_2^{-1}(M_{U})|}\over {\alpha_2^{-1}(M_{U})}},~
\Delta_{\pm} ~=~
 {{|\alpha_1^{-1}(M_{U\pm}) -
 \alpha_2^{-1}(M_{U\pm})|}\over {\alpha_2^{-1}(M_{U\pm})}}~.~\,
\label{D-Delta}
\end{eqnarray}

\begin{figure}[htb]
\centering
\includegraphics[width=8cm]{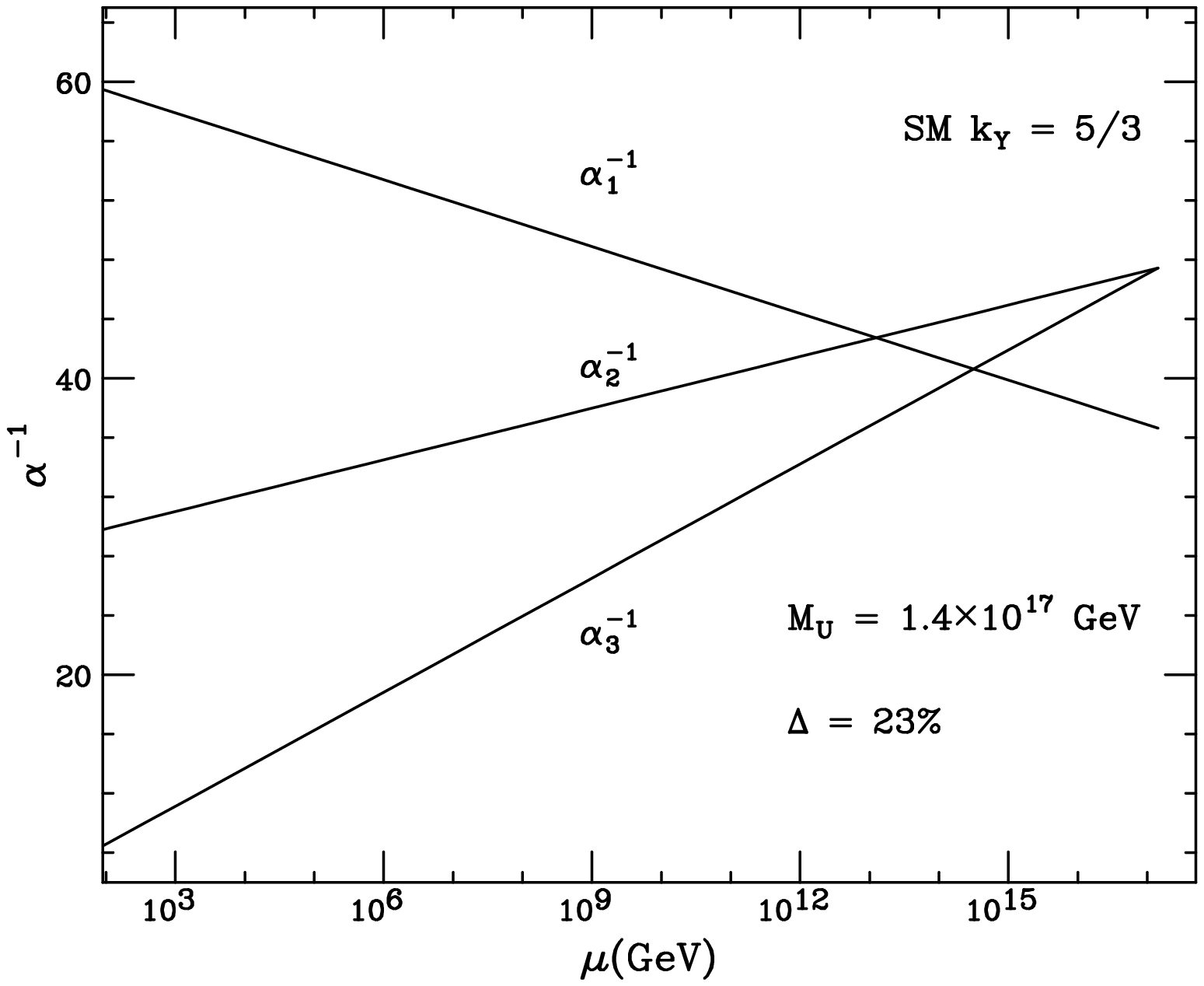}
\includegraphics[width=8cm]{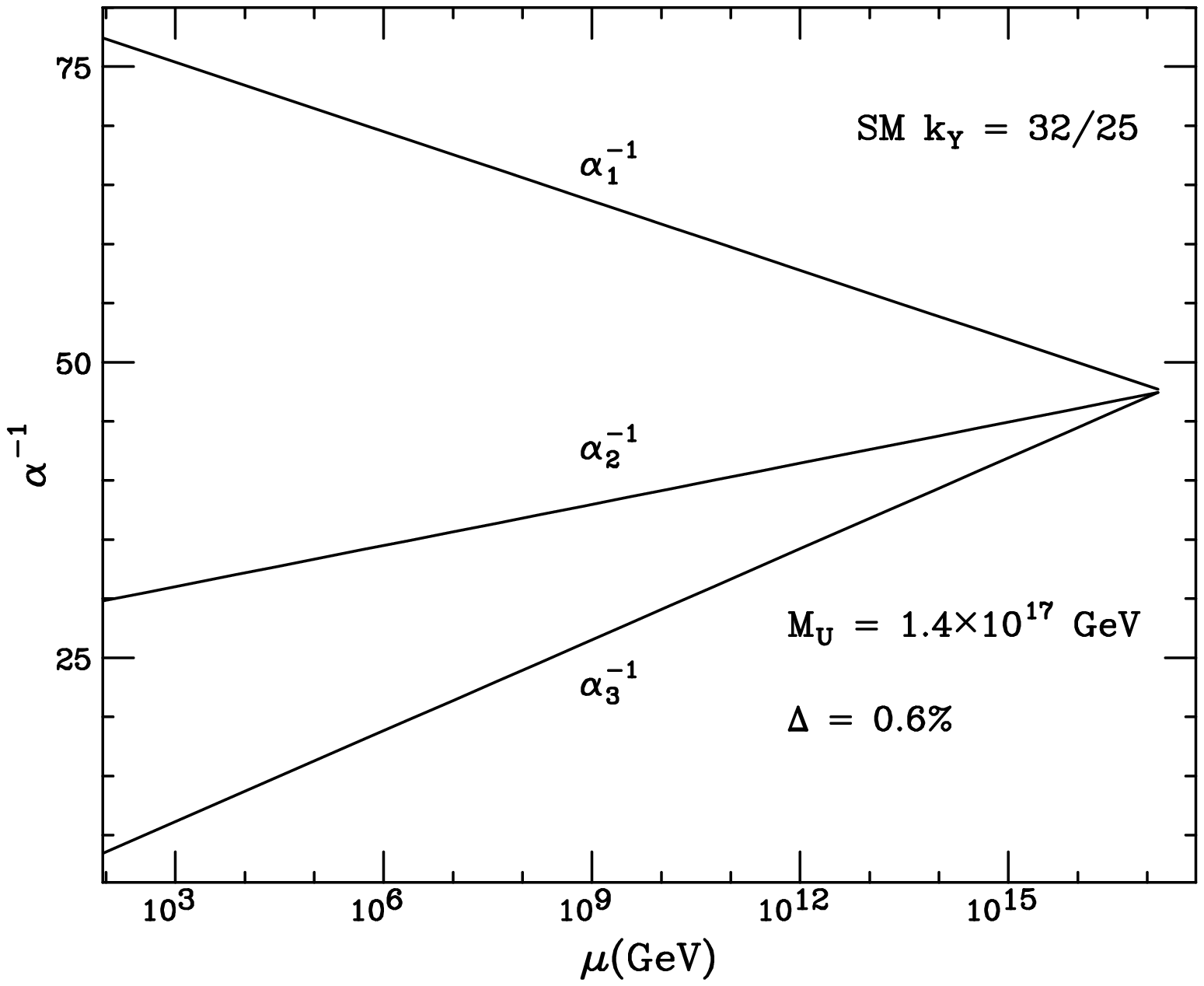}
\caption{One-loop gauge coupling unification for the SM with $k_Y=5/3$
(left) and $k_Y=32/25$ (right).}
\label{fig:1loop}
\end{figure}

In Fig.~\ref{fig:1loop}, with the central value of $\alpha_3$,
we show the one-loop RGE running of the SM
for canonical $U(1)_Y$ normalization $k_Y = 5/3$, 
and as a comparison, the results for $k_Y =32/25$.  
From the figures we see that the unification for $k_Y = 32/25$ is much better
than for $k_Y = 5/3$.   The convergences of the gauge couplings in the above
scenarios are summarized quantitatively in
Table~\ref{tbl:a3}, in which we list the unification scales $M_U$'s, and the relative
differences $\Delta$'s, as well as the values of $\alpha_2^{-1}(M_U)$ for the
central value of $\alpha_3(M_Z)$.  We confirm that the SM with canonical
normalization $k_Y=5/3$ is far from a good unification.  Introducing
supersymmetry significantly improves the convergence.  Meanwhile, the
same level of convergence can be achieved in all of the
non-supersymmetric models and in the MSSM I and MSSM II with
$k_Y=7/4$. In particular, the SM with $k_Y=32/25$ and
the 2HD SM with $k_Y=4/3$ have very good gauge coupling unification.

\begin{table}[htb]
\begin{center}
\begin{tabular}{|c|c|ccc|c|ccc|}
\hline
Model &$k_Y$ & $M_{U-}$ & $M_{U}$ & $M_{U+}$ & $\alpha_2^{-1}(M_U)$ &
$\Delta_-$ & $\Delta$ & $\Delta_+$ \\
\hline
SM      &  4/3  & 1.9  & 1.4  & 1.0  & 47.4 & 4.3  & 3.5  & 2.6 \\
SM      &  5/4  &      &      &      &      & 2.1  & 3.0  & 3.9 \\
SM      & 32/25 &      &      &      &      & 0.32 & 0.60 & 1.5 \\
SM      &  5/3  &      &      &      &      & 23.4 & 22.8 & 22.1 \\
2HD SM  &  4/3  & 0.45 & 0.33 & 0.24 & 45.8 & 0.25 & 1.1  & 2.0 \\
\hline
MSSM I  &  5/3  & 0.47 & 0.35 & 0.26 & 25.2 & 3.4  & 2.3  & 1.2 \\
MSSM II &  5/3  & 0.44 & 0.32 & 0.24 & 24.1 & 1.3  & 0.17 & 1.0 \\
MSSM I  &  7/4  & 0.47 & 0.35 & 0.26 & 25.2 & 8.0  & 7.0  & 5.9 \\
MSSM II &  7/4  & 0.44 & 0.32 & 0.24 & 24.1 & 6.0  & 4.9  & 3.8 \\
\hline
\end{tabular}
\end{center}
\caption{Convergences of the gauge couplings at one loop.
  The scale  $M_{U}$'s are in 
units of $10^{17}$ GeV, and the relative difference $\Delta$'s,
which are defined in Eq. (\ref{D-Delta}), are
percentile.}
\label{tbl:a3}
\end{table}

\begin{figure}[htb]
\centering
\includegraphics[width=8cm]{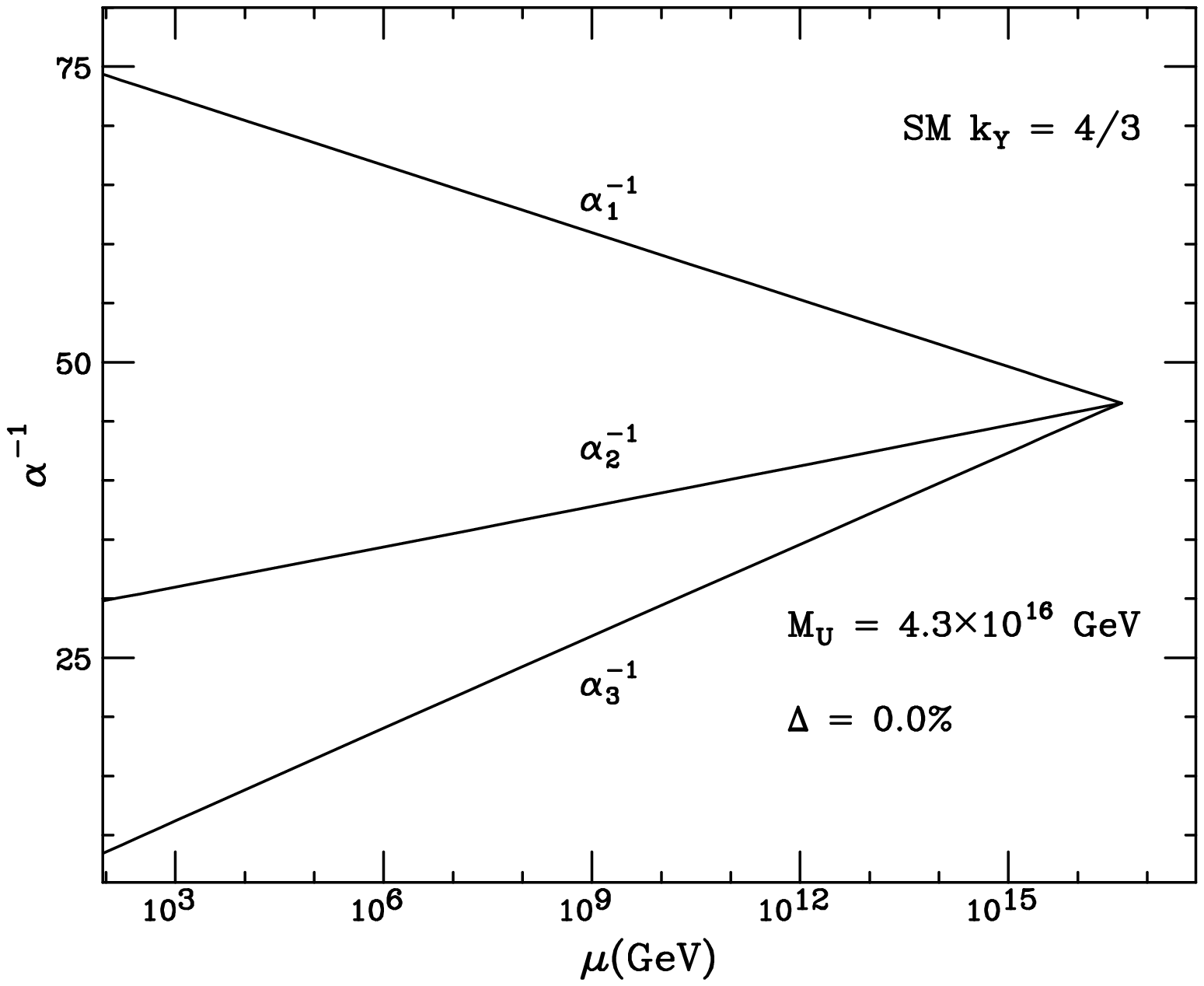}
\includegraphics[width=8cm]{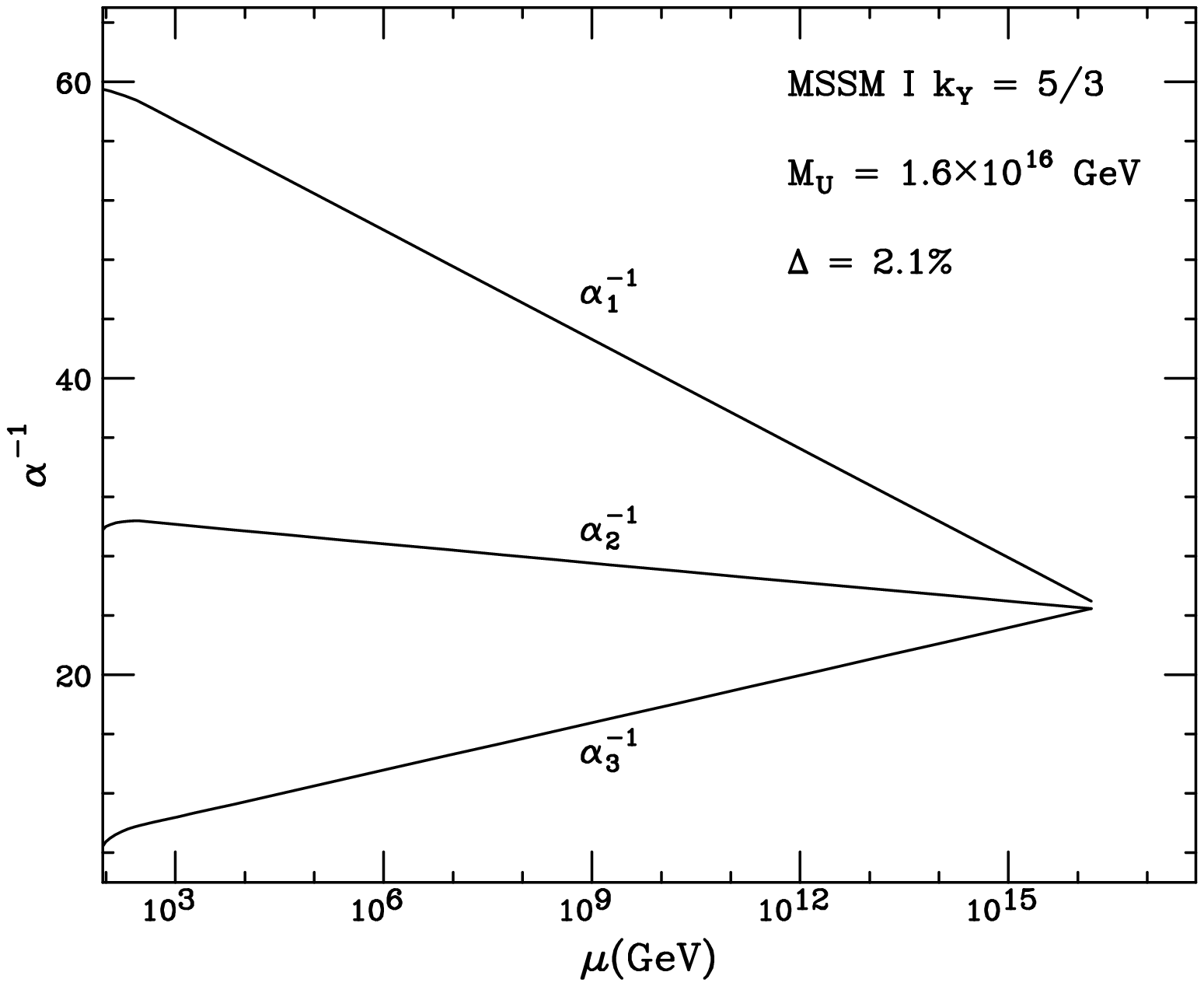}
\caption{Two-loop gauge coupling unification for the SM with $k_Y=4/3$
(left) and the MSSM I with $k_Y = 5/3$ (right). Note that $g_2$ is 
asymptotically free in the SM but not in the MSSM.}
\label{fig:2loop}
\end{figure}

To make a more precise evaluation of unification, it is necessary to
study two-loop RGE running.  We use the two-loop RGE running for
the gauge couplings and one-loop for the Yukawa 
couplings \cite{mac,Cvetic:1998uw,Barger:1992ac,Barger:1993gh,Martin:1993zk}.
The RGEs for different scenarios can be derived from the general expressions in
 Appendix~\ref{apdx}. 
The one-loop running dominates the evolution of the gauge couplings, 
with small corrections induced by the two-loop gauge coupling evolution and 
the one-loop Yukawa coupling evolution.  
With the central value of $\alpha_3$, we show
the gauge coupling unification for the SM with $k_Y = 4/3$ and the
MSSM I with $k_Y = 5/3$ in Fig.~\ref{fig:2loop}. 
 For the SM with $k_Y= 4/3$,  the
value of $\alpha_1$ precisely agrees with those of $\alpha_2$ and
$\alpha_3$ at the unification scale of $4.3 \times 10^{16}$ GeV. 
 On the other hand, for the scenario MSSM I with $k_Y =
5/3$, the value of $\alpha_1$ at $M_U = 1.6 \times 10^{16}$ GeV is
about $2.1\%$ higher than those of $\alpha_2$ and $\alpha_3$.  
 The unified
coupling strength in the SM is about one half of that in the supersymmetric 
models.  Table~\ref{tbl:2loop} shows the unification scales and the
relative differences for different scenarios. In comparison to
Table~\ref{tbl:a3}, we see that the two-loop corrections cause
$\alpha_2$ and $\alpha_3$ to unify at a smaller scale.  The two-loop
running improves the unification for the SM with $k_Y = 4/3$, but
worsens the unification for $k_Y = 32/25$.  The level of unification for
the MSSM I with $k_Y = 5/3$ remains the same.

\begin{table}[htb]
\begin{center}
\begin{tabular}{|c|c|ccc|c|ccc|}
\hline
Model &$k_Y$ & $M_{U-}$ & $M_{U}$ & $M_{U+}$ & $\alpha_2^{-1}(M_U)$ &
$\Delta_-$ & $\Delta$ & $\Delta_+$ \\
\hline
SM      & 4/3   & 0.31 & 0.43 & 0.57 & 46.6 & 0.87 & 0.00 & 0.85 \\
SM      & 5/4   &      &      &      &      & 7.6  & 6.6  & 5.7 \\
SM      & 32/25 &      &      &      &      & 5.0  & 4.1  & 3.2 \\
\hline
MSSM I  & 5/3   & 0.12 & 0.16 & 0.21 & 24.5 & 3.2  & 2.1  & 1.1 \\
MSSM II & 5/3   & 0.13 & 0.17 & 0.22 & 23.2 & 4.8  & 3.8  & 2.8 \\
MSSM I  & 7/4   & 0.12 & 0.16 & 0.21 & 24.5 & 1.9  & 2.9  & 3.8 \\
MSSM II & 7/4   & 0.13 & 0.17 & 0.22 & 23.2 & 0.3  & 1.3  & 2.2 \\
\hline
\end{tabular}
\end{center}
\caption{Convergences of the gauge couplings at two loop.
  The scale  $M_{U}$'s are in 
units of $10^{17}$ GeV, and the relative difference $\Delta$'s,
which are defined in Eq. (\ref{D-Delta}), are
percentile.}
\label{tbl:2loop}
\end{table}

\subsection{Comments on the $U(1)_Y$ Normalization in the MSSM}

\begin{figure}[htb]
\centering
\includegraphics[width=8cm]{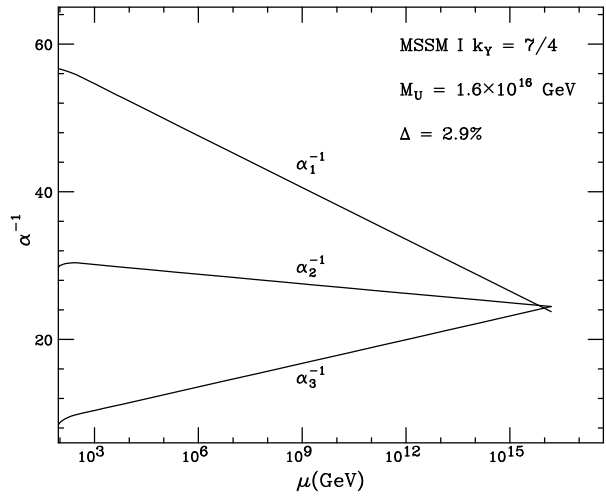}
\caption{Two-loop gauge coupling unification in the MSSM I with $k_Y=7/4$.}
\label{fig:2loop7o4}
\end{figure}

We want to further emphasize that, even within the MSSM, the canonical
$U(1)_Y$ normalization is not the only choice of normalization that produces
gauge coupling unification.  This means that a confirmation of the MSSM
from the discovery of supersymmetric particles at the LHC does not
necessarily imply that $k_Y = 5/3$.  In Tables~\ref{tbl:a3} and \ref{tbl:2loop},
 we show that the MSSM with $k_Y = 7/4$ can
produce the same level of unification as that in the MSSM with
canonical normalization $k_Y = 5/3$.
In Fig.~\ref{fig:2loop7o4}, we show the 
two-loop gauge coupling unification in the MSSM I with $k_Y=7/4$,
which is as good as the  MSSM I with $k_Y=5/3$.

\subsection{Higgs Boson Mass}

\begin{figure}[htb]
\centering
\includegraphics[width=8cm]{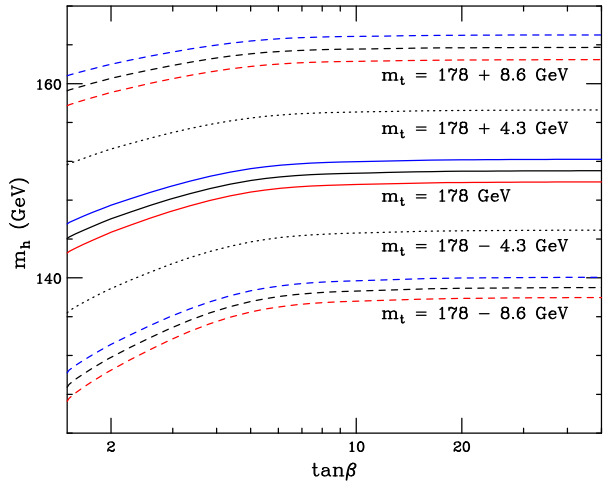}
\caption{The predicted Higgs
mass for the SM with $k_Y=4/3$. The red (lower) curves are for
$\alpha_3 + \delta\alpha_3$, the blue (upper) $\alpha_3 -
\delta\alpha_3$, and the black $\alpha_3$.  The dotted curves are for
$m_t \pm \delta m_t$, the dash ones for $m_t \pm 2 \delta m_t$, and
the solid ones for $m_t$.}
\label{fig:2loopm}
\end{figure}

If the Higgs particle is the only new physics discovered at the LHC
and the SM is thus confirmed as the low energy effective theory, the
most interesting parameter is the Higgs mass.  To be consistent with
string theory or quantum gravity, it is natural to have supersymmetry
in the fundamental theory. In supersymmetric models there
generically exist one pair of Higgs doublets $H_u$ and $H_d$.  We
define the SM Higgs doublet $H$, which is fine-tuned to have a small
mass, as $H \equiv -\cos\beta i \sigma_2 H_d^*+\sin\beta H_u$,
where $\sigma_2$ is the second Pauli matrix and $\tan\beta $ is a
mixing parameter~\cite{NASD,Barger:2004sf}.  For simplicity, we assume
that supersymmetry is broken at the GUT scale, {\it i.e.}, the
gauginos, squarks, sleptons, Higgsinos, and the other combination of
the scalar Higgs doublets ($\sin\beta i \sigma_2 H_d^*+\cos\beta H_u
$) have a universal supersymmetry breaking soft mass around the GUT
scale.  We can calculate the Higgs boson quartic coupling $\lambda$ at
the GUT scale \cite{NASD,Barger:2004sf}
\begin{equation}
\lambda({M_U}) = \frac{k_Y g_2^2(M_U) + g_1^2(M_U)}{4
k_Y} \cos^2 2\beta~,
\end{equation}
and then evolve it down to the weak scale. The renormalization group
equation for the quartic coupling is given in the Appendix~\ref{apdx}.
Using the one-loop effective Higgs potential with top quark radiative
corrections, we calculate the Higgs boson mass by minimizing the
effective potential
\be
V_{eff} = m_h^2 H^\dagger H - \frac{\lambda}{2!} (H^\dagger H)^2 -
\frac{3}{16\pi^2} h_t^4 (H^\dagger H)^2 \left[\log\frac{h_t^2 (H^\dagger
H)}{Q^2} - \frac{3}{2}\right]\,,
\ee
where $m_h^2$ is the Higgs mass square,
$h_t$ is the top quark Yukawa coupling, and the scale $Q$ is
chosen to be at the Higgs boson mass. 
For the ${\overline{MS}}$ top quark Yukawa coupling, we use the one-loop
corrected value~\cite{Arason:1991ic}, which is related to the top quark pole
mass by
\be
m_t = h_t v \left(1 + \frac{16}{3}\frac{g_3^2}{16\pi^2} - 2\frac{h_t^2}{16\pi^2}\right)\,.
\ee
For the SM with $k_Y=4/3$, the Higgs boson mass is shown as 
a function of $\tan\beta$ for different $m_t$ and
$\alpha_3$  in Fig.~\ref{fig:2loopm}.  If we vary
$\alpha_3$ within its $1\sigma$ range, $m_t$ within its $1\sigma$ and
$2\sigma$ ranges and $\tan\beta$ from $1.5$ to $50$, the predicted
Higgs boson mass will range from $127$ GeV to $165$ GeV.  A large part
of this uncertainty is due to the present uncertainty in the top quark
mass.  It is expected that
the top quark mass can be measured to about $1$ GeV accuracy at
the LHC~\cite{Beneke:2000hk}.  Assuming this accuracy and a central
value of $178$ GeV, the Higgs boson mass is predicted to be between
$141$ GeV and $154$ GeV.

Furthermore, for the SM with $k_Y=$ 5/4 and 32/25, the gauge coupling
unifications at two loop are similar to, but not as good as, that of the
SM with $k_Y=4/3$.  Following the same procedure as above, the
Higgs mass ranges for $k_Y=$ 5/4 and 32/25 turn out to be again from
$127$ GeV to $165$ GeV for the $2\sigma$ range of the top quark mass
and $1\sigma$ range of $\alpha_3$.
The Higgs mass ranges corresponding to the more precise (projected)
top quark mass
$m_t=178\pm1$ GeV are still between $141$ GeV and $154$ GeV.

\section{7D Orbifold GUTs}

In string model building, orbifold GUTs and their deconstruction,
 and 4D GUTs with product gauge groups,
the normalization for the $U(1)_Y$ need not be canonical. As explicit examples,
 we construct the 7D $SU(6)$ model with $k_Y=4/3$
and the 7D $SU(7)$ models with $k_Y=5/4$ and 32/25 on the
space-time $M^4\times T^2/Z_6 \times S^1/Z_2$ where
 charge quantization can be realized simultaneously.
We also briefly discuss the 7D orbifold $SU(8)$ model
with $k_Y=7/4$ and charge quantization, and comment on
a non-canonical $U(1)_Y$ normalization due to the
brane localized gauge kinetic terms in orbifold GUTs.

\subsection{Gauge Symmetry Breaking on $T^2/Z_6 \times S^1/Z_2$ Orbifold}

The 7D orbifold gauge symmetry breakings have not been studied previously,
so we discuss them in detail here.
The orbifold gauge symmetry breakings on the 7D
space-time $M^4\times  (S^1/Z_2)^3$ are similar to those
on the 6-dimensional (6D) space-time $M^4\times  (S^1/Z_2)^2$~\cite{Orbifold}.
Thus, we consider the 7D space-time $M^4\times T^2/Z_6 \times S^1/Z_2$,
with coordinates $x^{\mu}$, ($\mu = 0, 1, 2, 3$),
$x^5$, $x^6$ and $x^7$. Because $T^2$ is homeomorphic to $S^1\times S^1$,
we assume that the radii for the circles along the 
$x^5$, $x^6$ and $x^7$ directions
are $R_1$, $R_2$, and $R'$, respectively.
For simplicity, we define a complex
coordinate $z$ for $T^2$ and a real coordinate $y$ for $S^1$
\begin{eqnarray}
z \equiv{1\over 2} \left(x^5 + i x^6\right)~,~ y \equiv x^7~.~\,
\end{eqnarray}
In the complex coordinate,
the torus $T^2$ can be defined by $C^1$ moduloing the
equivalent classes: 
\begin{eqnarray}
z \sim z+ \pi R_1 ~,~ z \sim z +  \pi R_2 e^{{\rm i}\theta} ~.~\,
\end{eqnarray}

To define the $T^2/Z_6$ orbifold, we require that $R_1=R_2\equiv R$
and $\theta = \pi/3$.
The $T^2/Z_6 \times S^1/Z_2$ orbifold is obtained from $T^2 \times S^1$ 
by moduloing the equivalent classes
\begin{eqnarray}
\Gamma_T:~~z \sim \omega  z~;~~~~~\Gamma_S:~~ y \sim -y~,~ \,
\end{eqnarray}
where $\omega =e^{{\rm i}\pi/3} $.  
The fixed points under the $Z_6\times Z_2$ symmetry are
$(z, y)=(0, 0) $ and $(0, \pi R')$.

Note that our convention is as follows.
Suppose $G$ is a Lie group and $H$ is a subgroup of $G$.
We denote the commutant of $H$ in $G$ as $\{G/H\}$, {\it i.~e.},
\begin{equation}
\{G/H\} \equiv \{g \in G|gh=hg, ~{\rm for ~any} ~ h \in H\}~.~\,
\end{equation}

The ${\cal N}=1$ supersymmetry in 7 dimensions has 16 supercharges and
 corresponds to a ${\cal N}=4$ supersymmetry in 4 dimensions;
thus, only the gauge multiplet can be introduced in the bulk.  This
multiplet can be decomposed under the 4D
 ${\cal N}=1$ supersymmetry into a vector
multiplet $V$ and three chiral multiplets $\Sigma_1$, $\Sigma_2$, and 
$\Sigma_3$ in the adjoint representation, where the fifth and sixth 
components of the gauge
field, $A_5$ and $A_6$, are contained in the lowest component of $\Sigma_1$,
and the seventh component of the gauge
field $A_7$ is contained in the lowest component of $\Sigma_2$.
The SM fermions can be on the 3-branes at the $Z_6\times Z_2$
fixed points, or on the 4-branes at the $Z_6$ fixed points.

For the bulk gauge group $G$, we write down the  bulk action 
in the Wess-Zumino gauge and 4D ${\cal N}=1$ supersymmetry
language~\cite{NMASWS, NAHGW}
\begin{eqnarray}
  {\cal S} &=& \int d^7 x \Biggl\{
  {\rm Tr} \Biggl[ \int d^2\theta \left( \frac{1}{4 k g^2} 
  {\cal W}^\alpha {\cal W}_\alpha + \frac{1}{k g^2} 
  \left( \Sigma_3 \partial_z \Sigma_2 + \Sigma_1 \partial_y \Sigma_3
   - \frac{1}{\sqrt{2}} \Sigma_1 
  [\Sigma_2, \Sigma_3] \right) \right) 
\nonumber\\
  &&
+ {\rm H.C.} \Biggr] 
 + \int d^4\theta \frac{1}{k g^2} {\rm Tr} \Biggl[ 
  (\sqrt{2} \partial_z^\dagger + \Sigma_1^\dagger) e^{-V} 
  (-\sqrt{2} \partial_z + \Sigma_1) e^{V}
 + \partial_z^\dagger e^{-V} \partial_z e^{V}
\nonumber\\
  && + 
  (\sqrt{2} \partial_y + \Sigma_2^\dagger) e^{-V} 
  (-\sqrt{2} \partial_y + \Sigma_2) e^{V}
 + \partial_y e^{-V} \partial_y e^{V}
+ {\Sigma_3}^\dagger e^{-V} \Sigma_3 e^{V} 
\Biggr] \Biggr\}~.~\,
\label{eq:t2z6action}
\end{eqnarray}
From the above action, we obtain  
the transformations of the 4D vector multiplet and chiral multiplets 
\begin{eqnarray}
  V(x^{\mu}, ~\omega z, ~\omega^{-1} {\bar z},~y) &=& R_{\Gamma_T}
 V(x^{\mu}, ~z, ~{\bar z},~y) R_{\Gamma_T}^{-1}~,~\,
\label{TVtrans}
\end{eqnarray}
\begin{eqnarray}
  \Sigma_1(x^{\mu}, ~\omega z, ~\omega^{-1} {\bar z},~y) &=& 
\omega^{-1} R_{\Gamma_T}
\Sigma_1(x^{\mu}, ~z, ~{\bar z},~y) R_{\Gamma_T}^{-1}~,~\,
\label{T1trans}
\end{eqnarray}
\begin{eqnarray}
   \Sigma_2(x^{\mu}, ~\omega z, ~\omega^{-1} {\bar z},~y) &=& 
 R_{\Gamma_T} 
\Sigma_2(x^{\mu}, ~z, ~{\bar z},~y)  R_{\Gamma_T}^{-1}~,~\,
\label{T2trans}
\end{eqnarray}
\begin{eqnarray}
 \Sigma_3(x^{\mu}, ~\omega z, ~\omega^{-1} {\bar z},~y)  &=& 
\omega R_{\Gamma_T} 
\Sigma_3(x^{\mu}, ~z, ~{\bar z},~y) R_{\Gamma_T}^{-1}~,~\,
\label{T3trans}
\end{eqnarray}
\begin{eqnarray}
  V(x^{\mu}, ~z, ~ {\bar z},~-y) &=& R_{\Gamma_S}
 V(x^{\mu}, ~z, ~{\bar z},~y) R_{\Gamma_S}^{-1}~,~\,
\label{SVtrans}
\end{eqnarray}
\begin{eqnarray}
  \Sigma_1(x^{\mu}, ~ z, ~ {\bar z},~-y) &=& 
 R_{\Gamma_S}
\Sigma_1(x^{\mu}, ~z, ~{\bar z},~y) R_{\Gamma_S}^{-1}~,~\,
\label{S1trans}
\end{eqnarray}
\begin{eqnarray}
   \Sigma_2(x^{\mu}, ~ z, ~ {\bar z},~-y) &=& 
-  R_{\Gamma_S} 
\Sigma_2(x^{\mu}, ~z, ~{\bar z},~y)  R_{\Gamma_S}^{-1}~,~\,
\label{S2trans}
\end{eqnarray}
\begin{eqnarray}
 \Sigma_3(x^{\mu}, ~ z, ~ {\bar z},~-y)  &=& 
- R_{\Gamma_S} 
\Sigma_3(x^{\mu}, ~z, ~{\bar z},~y) R_{\Gamma_S}^{-1}~.~\,
\label{S3trans}
\end{eqnarray}
Here we introduced non-trivial
$R_{\Gamma_T}$ and $R_{\Gamma_S}$ to break the bulk gauge group $G$.

\subsection{$SU(6)$ Model with $k_Y=4/3$ }

First, let us consider the $SU(6)$ model, which has $k_Y=4/3$.
To break the $SU(6)$ gauge symmetry, we choose
the following $6\times 6$ matrix representations for 
$R_{\Gamma_T}$ and $R_{\Gamma_S}$
\begin{eqnarray}
R_{\Gamma_T} &=& {\rm diag} \left(+1, +1, +1,
 \omega^{n_1}, \omega^{n_1}, \omega^{n_2} \right)~,~\,
\end{eqnarray}
\begin{eqnarray}
R_{\Gamma_S} &=& {\rm diag} \left(+1, +1, +1, +1, +1, -1 \right)~,~\,
\end{eqnarray}
where $n_1$ and $n_2$ are positive integers, and $n_1 \not= n_2$.
Then, we obtain
\begin{eqnarray}
 \{SU(6)/R_{\Gamma_T}\} ~=~ SU(3)_C\times SU(2)_L\times U(1)_Y \times U(1)' ~,~\,
\end{eqnarray}
\begin{eqnarray}
\{SU(6)/R_{\Gamma_S}\} ~=~ SU(5)\times U(1) ~,~\,
\end{eqnarray}
\begin{eqnarray}
 \{SU(6)/\{R_{\Gamma_T} \cup R_{\Gamma_S}\}\}
~=~ SU(3)_C\times SU(2)_L\times U(1)_Y \times U(1)'~.~\,
\end{eqnarray}
Therefore,  for the zero modes, the 7D 
${\cal N} = 1 $ supersymmetric $SU(6)$ gauge symmetry is broken 
down to the 4D ${\cal N}=1$ supersymmetric
$SU(3)_C\times SU(2)_L\times U(1)_Y \times U(1)'$ \cite{Li:2001tx}.

We define the generators for the $U(1)_Y$ and $U(1)'$ in the $SU(6)$ as
\begin{eqnarray}
T_{U(1)_{Y}} &\equiv&  
{\rm diag}\left( {1\over 3}, {1\over 3}, {1\over 3}, 
- {1\over 3}, - {1\over 3}, - {1\over 3} \right)~,~\,
\label{SU6-GU1Y}
\end{eqnarray}
\begin{eqnarray}
T_{U(1)'} &\equiv& {\rm diag}\left( 0, 0, 0, 
{1\over 2}, {1\over 2}, -1 \right)~,~\,
\label{SU6-GU1P}
\end{eqnarray}
Because ${\rm tr} [T_{U(1)_{Y}}^2]=2/3$, we obtain $k_Y=4/3$.

The $SU(6)$ adjoint representation $\mathbf{35}$
is decomposed under the 
$SU(3)_C\times SU(2)_L\times U(1)_Y \times U(1)'$ gauge symmetry as
\begin{equation}
\mathbf{35} = \left(
\begin{array}{ccc}
\mathbf{(8,1)}_{Q00} & \mathbf{(3, \bar 2)}_{Q12} 
& \mathbf{(3, 1)}_{Q13}  \\
 \mathbf{(\bar 3,  2)}_{Q21} & \mathbf{(1,3)}_{Q00}
& \mathbf{(1, 2)}_{Q23}  \\
\mathbf{(\bar 3, 1)}_{Q31} & \mathbf{(1, \bar 2)}_{Q32}
& \mathbf{(1, 1)}_{Q00} 
\end{array}
\right) +  \mathbf{(1,1)}_{Q00}~,~\,
\end{equation}
where the  $\mathbf{(1,1)}_{Q00}$ in the third
diagonal entry of the matrix and the last term $\mathbf{(1,1)}_{Q00}$ 
denote the gauge fields for the 
$U(1)_Y \times U(1)'$ gauge symmetry.
Moreover,  the subscripts $Qij$, which are anti-symmetric
($Qij=-Qji$), are the charges under
the $U(1)_Y \times U(1)'$  gauge symmetry
\begin{eqnarray}
 Q00~=~ (\mathbf{0}, \mathbf{0}) ~,~
 Q12=(\mathbf{2\over 3}, \mathbf{-{1\over 2}})~,~
Q13=(\mathbf{2\over 3}, \mathbf{1})~,~
 Q23~=~ (\mathbf{0}, \mathbf{3\over 2})~.~\,
\end{eqnarray}

The $Z_6\times Z_2$ transformation properties for the decomposed components
of $V$, $\Sigma_1$, $\Sigma_2$,  and $\Sigma_3$ are
\begin{equation}
V : \left(
\begin{array}{ccc}
(1, +) & (\omega^{-n_1}, +) & 
(\omega^{-n_2}, -)  \\
(\omega^{n_1}, +) & (1, +) &  (\omega^{n_1-n_2}, -) \\
(\omega^{n_2}, -) & (\omega^{n_2-n_1}, -) 
& (1, +)  
\end{array}
\right)  +  (1, +) ~,~\,
\label{SU6trans-1}
\end{equation}
\begin{equation}
\Sigma_1 : \left(
\begin{array}{ccc}
(\omega^{-1}, +) & (\omega^{-n_1-1}, +) & 
(\omega^{-n_2-1}, -)  \\
(\omega^{n_1-1}, +) & (\omega^{-1}, +) &  (\omega^{n_1-n_2-1}, -) \\
(\omega^{n_2-1}, -) & (\omega^{n_2-n_1-1}, -) 
& (\omega^{-1}, +)  
\end{array}
\right) +  (\omega^{-1}, +) ~,~\,
\label{SU6trans-2}
\end{equation}
\begin{equation}
\Sigma_2 : \left(
\begin{array}{ccc}
(1, -) & (\omega^{-n_1}, -) & (\omega^{-n_2}, +)  \\
(\omega^{n_1}, -) & (1, -) &  (\omega^{n_1-n_2}, +) \\
(\omega^{n_2}, +) & (\omega^{n_2-n_1}, +) & (1, -) 
\end{array}
\right) +  (1, -) ~,~\,
\label{SU6trans-3}
\end{equation}
\begin{equation}
\Sigma_3 : \left(
\begin{array}{ccc}
(\omega, -) & (\omega^{-n_1+1}, -) & (\omega^{-n_2+1}, +) \\
(\omega^{n_1+1}, -) & (\omega, -) &  (\omega^{n_1-n_2+1}, +) \\
(\omega^{n_2+1}, +) & (\omega^{n_2-n_1+1}, +) &  (\omega, -) 
\end{array}
\right) +  (\omega, -) ~,~\,
\label{SU6trans-4}
\end{equation}
where the zero modes transform as $(1,+)$.
We choose
\begin{eqnarray}
n_1~=~2~,~~n_2~=~5~.~\,
\end{eqnarray}
From Eqs. (\ref{SU6trans-1})-(\ref{SU6trans-4}),
we obtain that, for the zero modes, the 7D
${\cal N} = 1 $ supersymmetric $SU(6)$ gauge symmetry is broken 
down to the 4D ${\cal N}=1$ supersymmetric
$SU(3)_C\times SU(2)_L\times  U(1)_Y \times  U(1)' $ gauge symmetry.
Also, we have no zero modes from chiral multiplets $\Sigma_1$ and $\Sigma_2$,
and we have one and only one zero mode from $\Sigma_3$ with
quantum number $\mathbf{(\bar 3, 1)}_{Q31}$ under the
$SU(3)_C\times SU(2)_L\times  U(1)_Y \times  U(1)' $ gauge symmetry,
which can be considered as the right-handed top quark
because its hypercharge is $\mathbf{-2/3}$.

On the 3-brane at the $Z_6\times Z_2$ fixed point
 $(z, y)= (0, 0)$, the preserved gauge symmetry 
is $SU(3)_C\times SU(2)_L\times U(1)_Y \times U(1)'$ \cite{Li:2001tx}.
Thus, on the observable 3-brane at $(z, y)= (0, 0)$, 
we can introduce  one pair of Higgs doublets and
three families of the SM quarks and leptons except
the right-handed top quark. Because the $U(1)_Y$ charge 
for the right-handed top quark is determined from the construction,
charge quantization can be achieved from the anomaly free
conditions and the gauge invariance
of the Yukawa couplings on the observable 3-brane.
Moreover, the $U(1)'$ anomalies can be cancelled by
assigning suitable $U(1)'$ charges to the SM 
quarks and leptons, for  example,
we assign the $U(1)'$ charges for the first, second
and third families of the SM fermions as
$\mathbf{+1}$, $\mathbf{0}$, $\mathbf{-1}$, respectively.
Also, the $U(1)'$ gauge symmetry can be broken at the GUT
scale by introducing
one pair of SM singlets with $U(1)'$ charges $\mathbf{\pm1}$ 
on the observable 3-brane.
Interestingly, this $U(1)'$ gauge symmetry may be considered
as a flavour symmetry, and then the SM fermion masses and
mixings may be explained naturally via the Froggatt-Nielsen mechanism \cite{FN}.
Furthermore, supersymmetry can be broken around the 
compactification scale, which can be considered as the GUT scale, 
for example, by the Scherk--Schwarz mechanism \cite{Scherk:1978ta}.

\subsection{$SU(7)$ Models with $k_Y=5/4$ and $32/25$ }

We will construct the $SU(7)$ models with $k_Y=5/4$ and $32/25$.
Because these two models are quite similar, we discuss them
simultaneously.

To break the $SU(7)$ gauge symmetry, we choose
the following $7\times 7$ matrix representations for 
$R_{\Gamma_T}$ and $R_{\Gamma_S}$
\begin{eqnarray}
R_{\Gamma_T} &=& {\rm diag} \left(+1, +1, +1,
 \omega^{n_1}, \omega^{n_1}, \omega^{n_1}, \omega^{n_2} \right)~,~\,
\end{eqnarray}
\begin{eqnarray}
R_{\Gamma_S} &=& {\rm diag} \left(+1, +1, +1, +1, +1, -1, -1 \right)~,~\,
\end{eqnarray}
where $n_1$ and $n_2$ are positive integers, and $n_1 \not= n_2$.
Then, we obtain
\begin{eqnarray}
 \{SU(7)/R_{\Gamma_T}\} ~=~ SU(3)_C\times SU(3)\times U(1) \times U(1)' ~,~\,
\end{eqnarray}
\begin{eqnarray}
\{SU(7)/R_{\Gamma_S}\} ~=~ SU(5)\times SU(2) \times U(1) ~,~\,
\end{eqnarray}
\begin{eqnarray}
 \{SU(7)/\{R_{\Gamma_T} \cup R_{\Gamma_S}\}\}
~=~ SU(3)_C\times SU(2)_L\times U(1)_Y \times U(1)_{\alpha} \times U(1)_{\beta}~.~\,
\end{eqnarray}
Therefore, we obtain that, for the zero modes, the 7D 
${\cal N} = 1 $ supersymmetric $SU(7)$ gauge symmetry is broken 
down to the 4D ${\cal N}=1$ supersymmetric
$SU(3)_C\times SU(2)_L\times U(1)_Y \times U(1)_{\alpha}
\times U(1)_{\beta}$ gauge symmetry  \cite{Li:2001tx}.

The $SU(7)$ adjoint representation $\mathbf{48}$
is decomposed under the 
$SU(3)_C\times SU(2)_L\times U(1)_Y \times U(1)_{\alpha}
\times U(1)_{\beta}$ gauge symmetry as
\begin{equation}
\mathbf{48} = \left(
\begin{array}{cccc}
\mathbf{(8,1)}_{Q00} & \mathbf{(3, \bar 2)}_{Q12} 
& \mathbf{(3, 1)}_{Q13} & \mathbf{(3,1)}_{Q14} \\
 \mathbf{(\bar 3,  2)}_{Q21} & \mathbf{(1,3)}_{Q00}
& \mathbf{(1, 2)}_{Q23} & \mathbf{(1, 2)}_{Q24} \\
\mathbf{(\bar 3, 1)}_{Q31} & \mathbf{(1, \bar 2)}_{Q32}
& \mathbf{(1, 1)}_{Q00} & \mathbf{(1, 1)}_{Q34} \\
\mathbf{(\bar 3, 1)}_{Q41} & \mathbf{(1, \bar 2)}_{Q42}
& \mathbf{(1, 1)}_{Q43} & \mathbf{(1, 1)}_{Q00} 
\end{array}
\right) +  \mathbf{(1,1)}_{Q00}~,~\,
\end{equation}
where the  $\mathbf{(1,1)}_{Q00}$ in the third and fourth 
diagonal entries of the matrix and the last term $\mathbf{(1,1)}_{Q00}$ 
denote the gauge fields for the $U(1)_Y \times U(1)_{\alpha}
\times U(1)_{\beta} $ gauge symmetry.
Moreover,  the subscripts $Qij$, which are anti-symmetric
($Qij=-Qji$), are the charges under
the $U(1)_Y \times U(1)_{\alpha} \times U(1)_{\beta}$ gauge symmetry.
The subscript $Q00~=~ (\mathbf{0}, \mathbf{0}, \mathbf{0})$, and 
the other subscripts $Qij$ with $i\not= j$ will be given for 
each model explicitly.

(1) $SU(7)$ model with  $k_Y=5/4$.
We define the generators for the 
$U(1)_Y \times U(1)_{\alpha} \times U(1)_{\beta}$ gauge symmetry
 as follows
\begin{eqnarray}
T_{U(1)_{Y}} &\equiv&  
{\rm diag}\left( {1\over 4}, {1\over 4}, {1\over 4},
 -{1\over 4}, -{1\over 4}, {1\over 4}, -{1\over 2}\right)~,~\,
\label{SU7-GU1Y}
\end{eqnarray}
\begin{eqnarray}
T_{U(1)_{\alpha}} &\equiv&  
{\rm diag}\left(1, 1, 1, -3, -3, -1, 4 \right)~,~\,
\label{SU7-GU1A}
\end{eqnarray}
\begin{eqnarray}
T_{U(1)_{\beta}} &\equiv&  
{\rm diag}\left( -{8\over 9}, -{8\over 9}, -{8\over 9},
-{1\over 2}, -{1\over 2}, 3, {2\over 3}\right)~.~\,
\label{SU7-GU1B}
\end{eqnarray}
Because ${\rm tr} [T_{U(1)_{Y}}^2]=5/8$, we obtain $k_Y=5/4$.
In this paper, we will choose convenient generators (normalizations) for
 $U(1)_{\alpha} \times U(1)_{\beta}$ gauge symmetry.

The $U(1)_Y \times U(1)_{\alpha} \times U(1)_{\beta}$ charges
$Qij$ are
\begin{eqnarray}
 Q12=(\mathbf{1\over 2}, \mathbf{4}, \mathbf{-{7\over {18}}})~,~
Q13=(\mathbf{0}, \mathbf{2}, \mathbf{-{{35}\over {9}}})~,~
Q14=(\mathbf{3\over 4}, \mathbf{-3}, \mathbf{-{{14}\over {9}}})~,~\,
\end{eqnarray}
\begin{eqnarray}
 Q23=(\mathbf{-{1\over 2}}, \mathbf{-2}, \mathbf{-{7\over {2}}})~,~
 Q24=(\mathbf{{1\over 4}}, \mathbf{-7}, \mathbf{-{7\over {6}}})~,~
Q34=(\mathbf{{3\over 4}}, \mathbf{-5}, \mathbf{{7\over {3}}})~.~\,
\end{eqnarray}

(2) $SU(7)$ model with $k_Y=32/25$.
We define the generators for the 
$U(1)_Y \times U(1)_{\alpha} \times U(1)_{\beta}$ gauge symmetry
 as follows
\begin{eqnarray}
T_{U(1)_{Y}} &\equiv&  
{\rm diag}\left( {3\over {10}}, {3\over {10}}, {3\over {10}},
-{2\over 5}, -{2\over 5}, {1\over {10}}, -{1\over 5}\right)~,~\,
\label{SU7-GU1Y-2}
\end{eqnarray}
\begin{eqnarray}
T_{U(1)_{\alpha}} &\equiv&  
{\rm diag}\left(1, 1, 1, 3, 3, -1, -8 \right)~,~\,
\label{SU7-GU1A-2}
\end{eqnarray}
\begin{eqnarray}
T_{U(1)_{\beta}} &\equiv&  
{\rm diag}\left(-47, -47, -47, -12, -12, 219, -54\right)~.~\,
\label{SU7-GU1B-2}
\end{eqnarray}
Because ${\rm tr} [T_{U(1)_{Y}}^2]=16/25$, we obtain $k_Y=32/25$.

The $U(1)_Y \times U(1)_{\alpha} \times U(1)_{\beta}$ charges
$Qij$ are
\begin{eqnarray}
 Q12=(\mathbf{7\over {10}}, \mathbf{-2}, \mathbf{-35})~,~
Q13=(\mathbf{1\over 5}, \mathbf{2}, \mathbf{-266})~,~
Q14=(\mathbf{1\over 2}, \mathbf{9}, \mathbf{7})~,~\,
\end{eqnarray}
\begin{eqnarray}
 Q23=(\mathbf{-{1\over 2}}, \mathbf{4}, \mathbf{-231})~,~
 Q24=(\mathbf{-{1\over 5}}, \mathbf{11}, \mathbf{42})~,~
Q34=(\mathbf{{3\over {10}}}, \mathbf{7}, \mathbf{273})~.~\,
\end{eqnarray}

The $Z_6\times Z_2$ transformation properties for the decomposed components
of $V$, $\Sigma_1$, $\Sigma_2$,  and $\Sigma_3$ are
\begin{equation}
V : \left(
\begin{array}{cccc}
(1, +) & (\omega^{-n_1}, +) & (\omega^{-n_1}, -) & 
(\omega^{-n_2}, -)  \\
(\omega^{n_1}, +) & (1, +) & (1, -) & (\omega^{n_1-n_2}, -) \\
(\omega^{n_1}, -) & (1, -) & (1, +) & (\omega^{n_1-n_2}, +) \\
(\omega^{n_2}, -) & (\omega^{n_2-n_1}, -) & (\omega^{n_2-n_1}, +) & (1, +)
\end{array}
\right)  +  (1, +) ~,~\,
\label{SU7-trans-1}
\end{equation}
\begin{equation}
\Sigma_1 : \left(
\begin{array}{cccc}
(\omega^{-1}, +) & (\omega^{-n_1-1}, +) & (\omega^{-n_1-1}, -) & 
(\omega^{-n_2-1}, -)  \\
(\omega^{n_1-1}, +) & (\omega^{-1}, +) & (\omega^{-1}, -) & (\omega^{n_1-n_2-1}, -) \\
(\omega^{n_1-1}, -) & (\omega^{-1}, -) & (\omega^{-1}, +) & (\omega^{n_1-n_2-1}, +) \\
(\omega^{n_2-1}, -) & (\omega^{n_2-n_1-1}, -) & (\omega^{n_2-n_1-1}, +)  & (\omega^{-1}, +)
\end{array}
\right) +  (\omega^{-1}, +) ~,~\,
\label{SU7-trans-2}
\end{equation}
\begin{equation}
\Sigma_2 : \left(
\begin{array}{ccccc}
(1, -) & (\omega^{-n_1}, -) & (\omega^{-n_1}, +) & 
(\omega^{-n_2}, +) \\
(\omega^{n_1}, -) & (1, -) & (1, +) & (\omega^{n_1-n_2}, +) \\
(\omega^{n_1}, +) & (1, +) & (1, -) & (\omega^{n_1-n_2}, -) \\
(\omega^{n_2}, +) & (\omega^{n_2-n_1}, +) & (\omega^{n_2-n_1}, -) & (1, -)
\end{array}
\right) +  (1, -) ~,~\,
\label{SU7-trans-3}
\end{equation}
\begin{equation}
\Sigma_3 : \left(
\begin{array}{ccccc}
(\omega, -) & (\omega^{-n_1+1}, -) & (\omega^{-n_1+1}, +) & 
(\omega^{-n_2+1}, +)  \\
(\omega^{n_1+1}, -) & (\omega, -) & (\omega, +) & (\omega^{n_1-n_2+1}, +) \\
(\omega^{n_1+1}, +) & (\omega, +) & (\omega, -) & (\omega^{n_1-n_2+1}, -) \\
(\omega^{n_2+1}, +) & (\omega^{n_2-n_1+1}, +) & (\omega^{n_2-n_1+1}, -) & (\omega, -)
\end{array}
\right) +  (\omega, -) ~,~\,
\label{SU7-trans-4}
\end{equation}
where the zero modes transform as $(1,+)$.
We choose
\begin{eqnarray}
n_1~=~2~,~~n_2~=~4~.~\,
\end{eqnarray}
From Eqs. (\ref{SU7-trans-1})-(\ref{SU7-trans-4}),
we obtain that, for the zero modes, the 7D 
${\cal N} = 1 $ supersymmetric $SU(7)$ gauge symmetry is broken 
down to the 4D ${\cal N}=1$ supersymmetric
$SU(3)_C\times SU(2)_L\times  U(1)_Y \times U(1)_{\alpha} \times U(1)_{\beta} $
 gauge symmetry. Also, we have no zero modes from chiral 
multiplets $\Sigma_1$ and $\Sigma_3$,
and we have  only one pair of zero modes from $\Sigma_2$ with
quantum numbers $\mathbf{(1,  2)}_{Q23}$ and $\mathbf{(1, \bar 2)}_{Q32}$
under the 
$SU(3)_C\times SU(2)_L\times  U(1)_Y \times U(1)_{\alpha} \times U(1)_{\beta} $
 gauge symmetry, which can be considered as one pair of the Higgs doublets
$H_d$ and $H_u$ in the supersymmetric models, respectively.

On the 3-brane at the 
 $(z, y)= (0, 0)$, the preserved gauge symmetry is 
$SU(3)_C\times SU(2)_L\times U(1)_Y \times 
U(1)_{\alpha} \times U(1)_{\beta} $ \cite{Li:2001tx}.
Thus, on the observable 3-brane at $(z, y)= (0, 0)$, 
we can introduce three families of the SM fermions. 
Because the $U(1)_Y$ hypercharges for one pair of 
Higgs doublets $H_d$ and $H_u$ are determined from the model building,
charge quantization can be achieved from the anomaly free
conditions and the gauge invariance
of the Yukawa couplings on the observable 3-brane.
Moreover, because $H_d$ and $H_u$ are vector-like under
the $SU(3)_C\times SU(2)_L\times U(1)_Y \times U(1)_{\alpha} 
\times U(1)_{\beta} $ gauge symmetry, there are no 
$U(1)_{\alpha}$ and $U(1)_{\beta} $ anomalies from them.
The $U(1)_{\alpha} \times U(1)_{\beta} $ gauge symmetry can 
be broken at the GUT scale by introducing
two pairs of the SM singlets with non-trivial
 $U(1)_{\alpha} \times U(1)_{\beta} $ charges on the observable 3-brane.
The remarks at the end of above subsection for the fermion
spectrum and supersymmetry breaking apply here as well.

\subsection{$SU(8)$ Model with $k_Y=7/4$ }

We briefly present a 7D orbifold $SU(8)$ model with $k_Y=7/4$,
which gives an alternative $U(1)_Y$ normalization for the MSSM.
To avoid confusion, we emphasize that in this model we consider
TeV-scale supersymmetry breaking.
To break the $SU(8)$ gauge symmetry, we choose
the following $8\times 8$ matrix representations for 
$R_{\Gamma_T}$ and $R_{\Gamma_S}$
\begin{eqnarray}
R_{\Gamma_T} &=& {\rm diag} \left( +1, +1, +1,
 \omega^{n_1}, \omega^{n_1}, \omega^{n_1}, \omega^{n_2}, \omega^{n_2} \right)~,~\,
\end{eqnarray}
\begin{eqnarray}
R_{\Gamma_S} &=& {\rm diag} \left(+1, +1, +1, +1, +1, -1, -1, +1\right)~,~\,
\end{eqnarray}
where $n_1$ and $n_2$ are positive integers, and $n_1 \not= n_2$.
Then, we find
\begin{eqnarray}
 \{SU(8)/R_{\Gamma_T}\} ~=~ SU(3)_C\times SU(3)\times SU(2) \times U(1)^2 ~,~\,
\end{eqnarray}
\begin{eqnarray}
\{SU(8)/R_{\Gamma_S}\} ~=~ SU(6)\times SU(2) \times U(1) ~,~\,
\end{eqnarray}
\begin{eqnarray}
 \{SU(8)/\{R_{\Gamma_T} \cup R_{\Gamma_S}\}\}
~=~SU(3)_C \times SU(2)_L\times U(1)_Y \times U(1)_{\alpha}
\times U(1)_{\beta} \times U(1)_{\gamma}~.~\,
\end{eqnarray}
Therefore, we obtain that, for the zero modes, the 7D
${\cal N} = 1 $ supersymmetric $SU(8)$ gauge symmetry is broken 
down to the 4D ${\cal N}=1$ supersymmetric
$SU(3)_C\times SU(2)_L\times U(1)_Y \times U(1)_{\alpha}
\times U(1)_{\beta} \times U(1)_{\gamma}$ gauge symmetry  \cite{Li:2001tx}.

We define the $U(1)_Y$ generator as following
\begin{eqnarray}
T_{U(1)_{Y}} &\equiv&  
{\rm diag}\left( {1\over 4}, {1\over 4}, {1\over 4},
 {1\over 4}, {1\over 4}, -{1\over 4}, -{1\over 2}, -{1\over 2}\right)~.~\,
\label{SU8-GU1Y}
\end{eqnarray}
Because ${\rm tr} [T_{U(1)_{Y}}^2]=7/8$, we obtain $k_Y=7/4$.
We choose
\begin{eqnarray}
n_1~=~5~,~~  n_2 ~=~4~.~\,
\end{eqnarray}
After detailed calculations, we find that
there are one pair of Higgs doublets $H_d$ and $H_u$
 from the zero modes of $\Sigma_2$ and some exotic particles from the zero modes
of the chiral multiplets $\Sigma_1$, $\Sigma_2$ and $\Sigma_3$.

Similar to the above subsection, we introduce  
three families of SM fermions on the observable 3-brane 
at $(z, y)= (0, 0)$ and charge quantization can be realized.
The $U(1)_{\alpha} \times U(1)_{\beta} \times U(1)_{\gamma}$ 
gauge symmetry can  be considered as a flavour symmetry,
and it can be broken at the GUT scale by introducing
three pairs of the SM singlets with non-trivial
 $U(1)_{\alpha} \times U(1)_{\beta} \times U(1)_{\gamma}$ charges
on the observable 3-brane. In addition,
the exotic particles from the zero modes
of  chiral multiplets $\Sigma_i$ can be made very 
heavy after the $U(1)_{\alpha} \times U(1)_{\beta} \times U(1)_{\gamma}$ 
gauge symmetry breaking by coupling them to the extra fields on
the observable 3-brane.

\subsection{Remarks on Another Possibility}

In 5-dimensional (5D) orbifold $SU(5)$,
or 6D orbifold $SO(10)$~\cite{Orbifold}, 
the  SM gauge couplings $g_Y$, $g_2$, and $g_3$ 
at the unification scale are obtained from compactification, {\it i.e.},
\begin{eqnarray}
g_Y=g_Y^{B}~,~~g_2 = g_2^B~,~~g_3=g^B_3~,~\,
\end{eqnarray}
where $g_Y^{B}$, $g_2^B$ and $g^B_3$ are the properly normalized
4D effective gauge couplings from 5D $SU(5)$
or 6D $SO(10)$ gauge kinetic terms.
Because we have ${\sqrt {5\over 3}} g_Y^{B} = g_2^B = g^B_3$ at the
unification scale, we obtain
\begin{eqnarray}
{\sqrt {5\over 3}} g_Y = g_2 = g_3~.~\,
\end{eqnarray}

However, on the 3-branes at the fixed points, only the SM
or SM-like gauge symmetry should be preserved, so there
exists the possibility that one may
introduce the 3-brane localized gauge kinetic terms from
the effective field theory point of view~\cite{Dvali:2000rx}. Thus,
the effective SM gauge couplings $g_Y$, $g_2$, and $g_3$ 
at the unification scale become
\begin{eqnarray}
{1\over g_Y^2} ={1\over g_Y^{B2}} +{1\over g_Y^{\prime 2}}~,~~
{1\over g_2^2} ={1\over g_2^{B2}} +{1\over g_2^{\prime 2}}~,~~
{1\over g_3^2} ={1\over g_3^{B2}} +{1\over g_3^{\prime 2}}~,~\,
\end{eqnarray}
where  $g_Y^{\prime}$, $g_2^{\prime}$ and $g^{\prime}_3$ are the
 properly normalized 4D effective gauge couplings from
3-brane localized gauge kinetic terms. In general, we have
\begin{eqnarray}
{\sqrt {5\over 3}} g_Y^{\prime} \not= g_2^{\prime} \not= g_3^{\prime}~.~\,
\end{eqnarray}
Thus, at the unification scale, we obtain
\begin{eqnarray}
{\sqrt {5\over 3}} g_Y \not= g_2 \not= g_3~.~\,
\end{eqnarray}
Therefore, the $U(1)_Y$ (and other gauge factors) normalization is not canonical.

In this paper we just point out this possibility, but we do not
take it seriously for these reasons: (1) To achieve the
gauge coupling unification in the SM,
we need to fine-tune the brane localized gauge kinetic terms;
(2) There are no such brane localized gauge kinetic terms
in the orbifold compactifications of the weakly coupled
heterotic string theory~\cite{Dixon:1985jw}; thus, whether such terms
do exist is unresolved.

\section{Discussion and Conclusions}

How to test our models with different $U(1)_Y$ normalizations
is an interesting question. However, 
 it is very difficult for two reasons. First,
 there exist unkown threshhold corrections
(including the supersymmetric threshold corrections) close to the GUT scale
 because a lot of new particles may appear, and 
 higher-dimensional operators may also contribute to the gauge couplings,
so the concrete prediction for one of the three SM gauge couplings at
 the weak scale due to the RGE running from the unification
scale will be GUT model dependent.
 Furthermore, the RGE running of the gauge couplings in the SM
for different $U(1)_Y$ normalizations will not 
cause any physically different results at low energy, 
{\it i.~e.}, the SM
with different $U(1)_Y$ normalizations are equivalent as low
 energy effective theories.

The string landscape suggests that the supersymmetry breaking scale
can be high and then the simplest low energy effective theory is
just the SM. Considering GUT scale supersymmetry breaking,
we showed that gauge coupling unification in the SM can be
achieved at about $10^{16-17}$ GeV for $k_Y=4/3$, 5/4, 32/25.
Especially for $k_Y=4/3$, gauge coupling unification in the SM
is well satisfied at two loop order. 
We also predicted that the Higgs mass is in the range 127 GeV to 165 GeV
by varying $\alpha_3$ within its $1\sigma$ range, $m_t$ within its 
$2\sigma$ range and $\tan\beta$ from $1.5$ to $50$.
For a future top quark mass measurement of value and uncertainty
 $m_t=178\pm1$ GeV, for example, we obtained a Higgs boson mass 
between $141$ GeV and $154$ GeV. Moreover,
we  pointed out that gauge coupling unification in the MSSM 
does not necessarily imply $k_Y=5/3$. We showed that 
gauge coupling unification in the MSSM 
can  be achieved at the same level by choosing $k_Y=7/4$.

Furthermore, 
we constructed a 7D $SU(6)$ model with $k_Y=4/3$
and 7D $SU(7)$ models with $k_Y=5/4$ and 32/25
on the space-time $M^4\times T^2/Z_6 \times S^1/Z_2$.
In these models, the $SU(6)$ and $SU(7)$ gauge symmetries
can be broken down to the SM-like gauge symmetries via orbifold
projections and then broken further down to the SM gauge symmetry by
the Higgs mechanism. The right-handed top quark in the $SU(6)$ model
and one pair of the Higgs doublets in the $SU(7)$ models can
be obtained from the zero modes of the bulk vector multiplet, with their
hypercharges determined by the constructions.
Then charge quantization can be achieved from the anomaly free
conditions and the gauge invariance
of the Yukawa couplings. 
The extra $U(1)$ gauge symmetries can be considered
as  flavour symmetries, and then the SM fermion masses and
mixings may be explained naturally via the Froggatt-Nielsen mechanism \cite{FN}.
The supersymmetry can be broken at the GUT scale 
by the Scherk--Schwarz mechanism \cite{Scherk:1978ta}.
We also briefly presented a 7D orbifold $SU(8)$ model
with $k_Y=7/4$ and charge quantization
and commented on non-canonical $U(1)_Y$ normalization due to the
brane localized gauge kinetic terms in orbifold GUTs.

\section*{Acknowledgments}

 This research was supported by the U.S.~Department of Energy
under Grants No.~DE-FG02-95ER40896, DE-FG02-96ER40969 and DOE-EY-76-02-3071,
 by the National Science
Foundation under Grant No.~PHY-0070928, and by the University
of Wisconsin Research Committee with funds granted by the Wisconsin
Alumni Research Foundation.

\appendix
\section{ Renormalization Group Equations}
\label{apdx}
In this Appendix, following our convention in Ref.~\cite{Barger:2004sf},
we give the renormalization group equations in the
SM and supersymmetric models with a general normalization factor
$k_Y$.  The general formulae for the renormalization group equations
in the SM are given in Refs.~\cite{mac,Cvetic:1998uw}, and those for
the supersymmetric models are given in
Refs.~\cite{Barger:1992ac,Barger:1993gh,Martin:1993zk}.

First, we present the renormalization group equations in the SM.  The
two-loop renormalization group equations for the gauge couplings are
\begin{equation}
(4\pi)^2\frac{d}{dt}~ g_i=g_i^3b_i +\frac{g_i^3}{(4\pi)^2}
\left[ \sum_{j=1}^3 B_{ij}g_j^2-\sum_{\alpha=u,d,e} d_i^\alpha
{\rm Tr}\left( h^{\alpha \dagger}h^{\alpha}\right) \right] ~,~\,
\label{2lgauge}
\end{equation}
The beta-function coefficients are
\begin{eqnarray}
&&b=\left(\frac{41}{6} \frac{1}{k_Y},-\frac{19}{6},-7\right) ~,~
B=\pmatrix{\frac{199}{18} \frac{1}{k_Y^2} &
\frac{27}{6} \frac{1}{k_Y} &\frac{44}{3} \frac{1}{k_Y} \cr 
\frac{3}{2} \frac{1}{k_Y} & \frac{35}{6}&12 \cr
\frac{11}{6} \frac{1}{k_Y} &\frac{9}{2}&-26} ~,~\\
&&d^u=\left(\frac{17}{6} \frac{1}{k_Y} ,\frac{3}{2},2\right) ~,~
d^d=\left(\frac{5}{6} \frac{1}{k_Y},\frac{3}{2},2\right) ~,~
d^e=\left(\frac{5}{2} \frac{1}{k_Y},\frac{1}{2},0\right) ~.~\,
\end{eqnarray}

Since the contributions in Eq.~(\ref{2lgauge}) from the Yukawa
couplings arise from two-loop diagrams, we only need to include
Yukawa coupling evolution at one-loop order.  The one-loop
renormalization group equations for Yukawa couplings are
\begin{eqnarray}
(4\pi)^2\frac{d}{dt}~h^u&=&h^u\left( -\sum_{i=1}^3c_i^ug_i^2
+\frac{3}{2}
h^{u \dagger}h^{u}
-\frac{3}{2}
h^{d \dagger}h^{d}
+\Delta_2\right) ~,~\\
(4\pi)^2\frac{d}{dt}~h^d&=&h^d\left( -\sum_{i=1}^3c_i^dg_i^2
-\frac{3}{2}
h^{u \dagger}h^{u}
+\frac{3}{2}
h^{d \dagger}h^{d}
+\Delta_2 \right)~,~\\
(4\pi)^2\frac{d}{dt}~h^e&=&h^e\left( -\sum_{i=1}^3c_i^eg_i^2
+\frac{3}{2}
h^{e \dagger}h^{e}
+ \Delta_2 \right) ~,~\,
\label{SMY}
\end{eqnarray}
where 
\begin{eqnarray}
c^u=\left( \frac{17}{12} \frac{1}{k_Y}, \frac{9}{4}, 8\right) ~,~
c^d=\left( \frac{5}{12} \frac{1}{k_Y}, \frac{9}{4}, 8\right) ~,~
c^e=\left( \frac{15}{4} \frac{1}{k_Y}, \frac{9}{4}, 0\right) ~,~
\end{eqnarray}
\begin{eqnarray}
\Delta_2 &=& {\rm Tr} ( 3h^{u \dagger}h^{u}+3 h^{d \dagger}h^{d}+
h^{e \dagger}h^{e})  ~.~
\end{eqnarray}

The one-loop renormalization group equation for the Higgs quartic coupling is 
\begin{eqnarray}
(4\pi)^2\frac{d}{dt}~\lambda  &=&12 \lambda^2
-\left(3 \frac{1}{k_Y} g_1^2 + 9 g_2^2 \right) \lambda
+{9\over 4} \left( \frac{1}{3} \frac{1}{k_Y^2} g_1^4 
+ {2\over 3} \frac{1}{k_Y} g_1^2g_2^2 + g_2^4 \right)
\nonumber\\&&
+4\Delta_2 \lambda - 4 \Delta_4 ~,~\,
\end{eqnarray}
where
\begin{eqnarray}
\Delta_4 &=& {\rm Tr} \left[ 3 (h^{u \dagger}h^{u})^2+3 (h^{d \dagger}h^{d})^2
+ (h^{e \dagger}h^{e})^2\right]  ~.~
\end{eqnarray}

Second, we give the beta-function coefficients for supersymmetric
models.  The two-loop renormalization group equations 
for the gauge couplings are the same as Eq.~(\ref{2lgauge}).
The beta-function coefficients are modified due to the new particle
contents. They are
\begin{eqnarray}
&&b=\left(11 \frac{1}{k_Y},1,-3\right) ~,~ 
B=\pmatrix{\frac{199}{9}
\frac{1}{k_Y^2}& 9 \frac{1}{k_Y}&\frac{88}{3} \frac{1}{k_Y} \cr
3\frac{1}{k_Y} & 25&24 \cr
\frac{11}{3}\frac{1}{k_Y} & 9 & 14} ~,~ \\
&&d^u=\left(\frac{26}{3} \frac{1}{k_Y},6,4\right) ~,~
d^d=\left(\frac{14}{3}\frac{1}{k_Y},6,4\right) ~,~
d^e=\left(6 \frac{1}{k_Y},2,0\right) ~.~
\end{eqnarray}

The one-loop renormalization group equations for Yukawa couplings are
\begin{eqnarray}
(4\pi)^2\frac{d}{dt}~y^u&=& y^u
\left[ 3 y^{u \dagger} y^{u}+ y^{d \dagger} y^{d}
+3{\rm Tr}( y^{u \dagger} y^{u}) 
-\sum_{i=1}^3c_i^ug_i^2 \right]~,~\\
(4\pi)^2\frac{d}{dt}~y^d&=& y^d
\left[ y^{u \dagger} y^{u} + 3 y^{d \dagger} y^{d}
+{\rm Tr}(3 y^{d \dagger} y^{d}
+ y^{e \dagger} y^{e}) 
-\sum_{i=1}^3c_i^dg_i^2 \right]~,~\\
(4\pi)^2\frac{d}{dt}~y^e&=& y^e
\left[ 3 y^{e \dagger} y^{e}+{\rm Tr}(3 y^{d \dagger} y^{d}
+ y^{e \dagger} y^{e}) 
-\sum_{i=1}^3c_i^eg_i^2 \right] ~,~\\
\end{eqnarray}
where
\begin{eqnarray}
&& c^u=\left( \frac{13}{9} \frac{1}{k_Y}, 3, \frac{16}{3}\right) ~,~
c^d=\left( \frac{7}{9}\frac{1}{k_Y}, 3, \frac{16}{3}\right) ~,~
c^e=\left( 3\frac{1}{k_Y}, 3, 0\right) ~.~
\end{eqnarray}


\end{document}